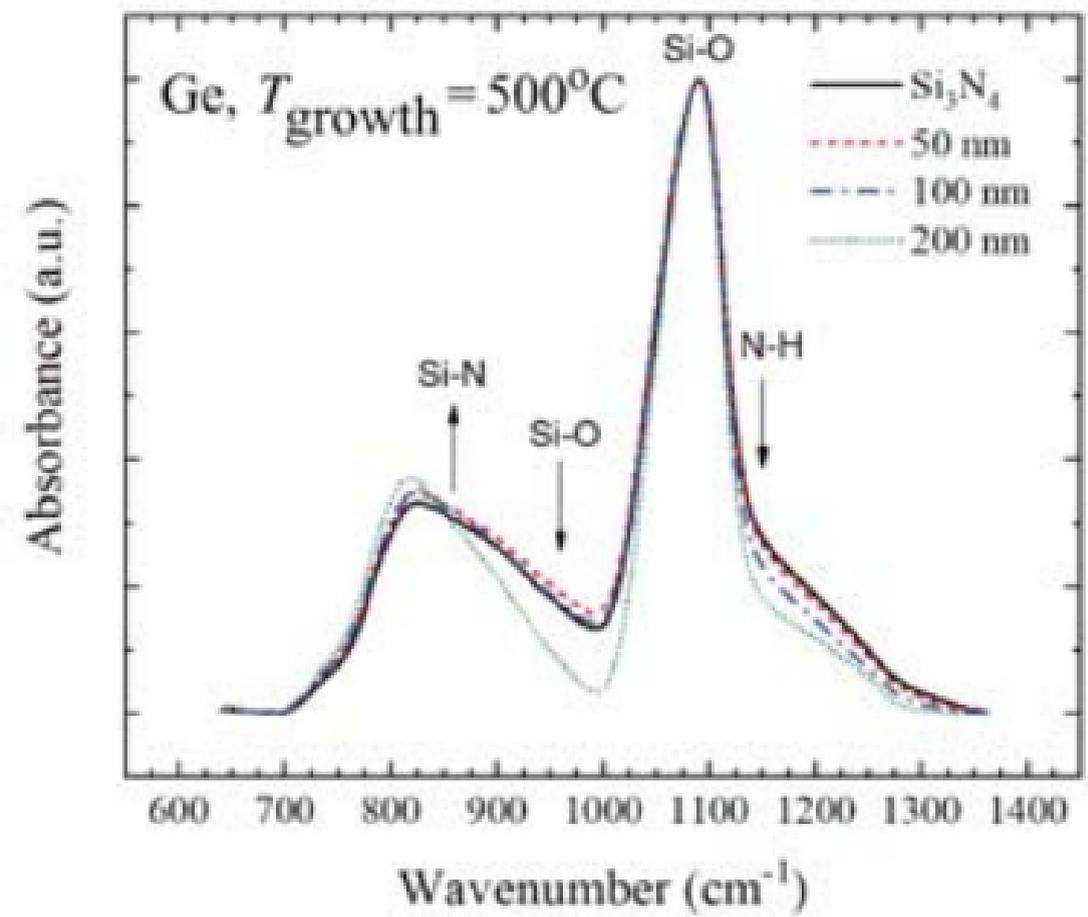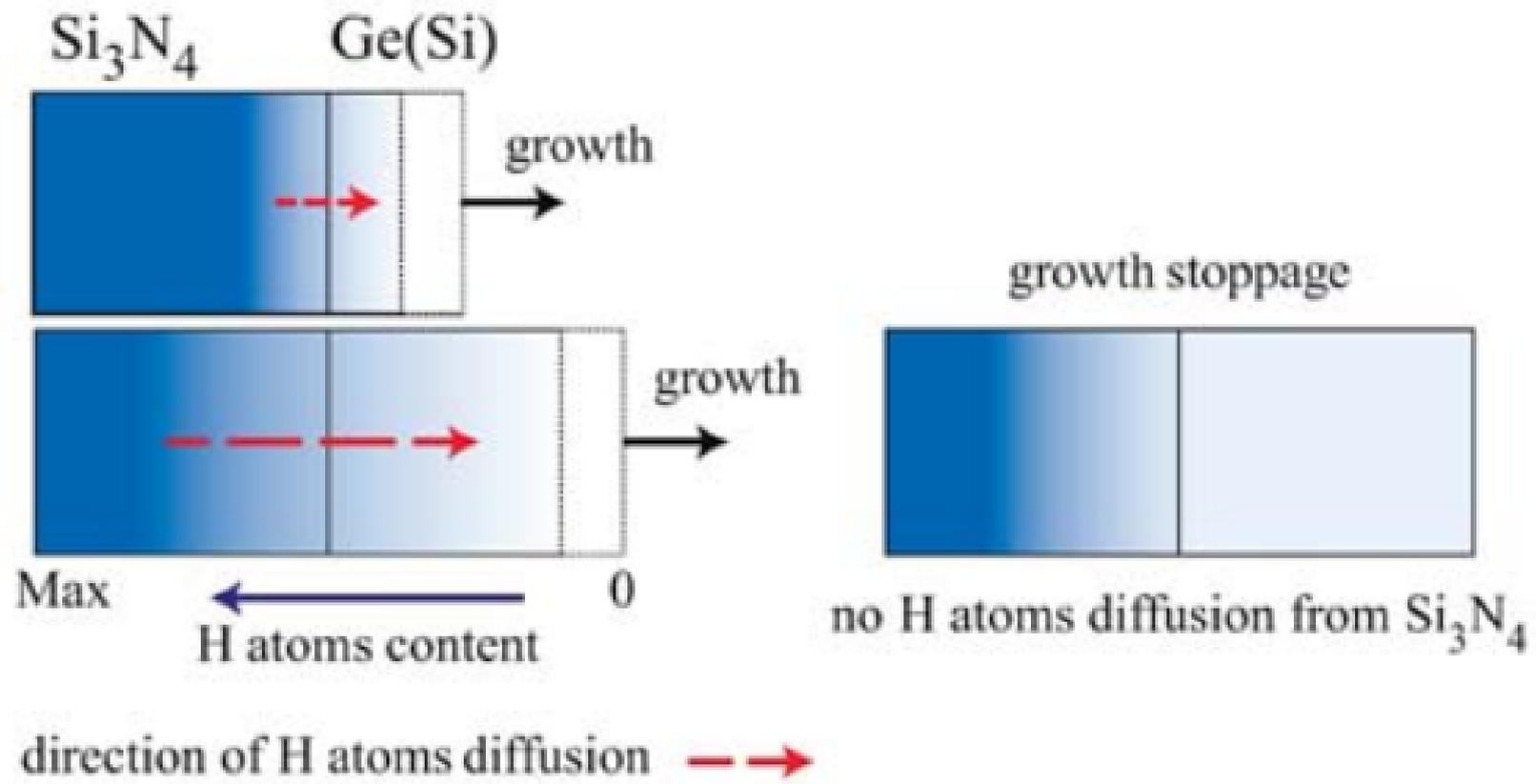

H atoms come off N–H bonds in $Si_3N_4$ to move into growing Ge(Si) films.
Ge(Si) layer thickness controls the diffusion processes in growing film.

Highlights

- Diffusion in Ge and Si films growing on $Si_3N_4/SiO_2/Si$ substrates is studied
- Ge (Si) grains coarsen with the increase in poly-Ge (Si) film thickness
- H atoms come off N–H bonds in $Si_3N_4$ to move into growing Ge (Si) films
- Ge (Si) atoms migrate into $Si_3N_4$ layer to replace H atoms via bonding to N
- Ge (Si) layer thickness controls the diffusion processes in growing film

# Influence of the thickness of Si and Ge films deposited on $Si_3N_4$/$SiO_2$/Si substrates on their structure and diffusion of hydrogen atoms from $Si_3N_4$ layers

Larisa V. Arapkina[1], Kirill V. Chizh[1], Dmitry B. Stavrovskii[1],

Alexey A. Klimenko[2], Alexander A. Dudin[2], Vladimir P. Dubkov[1],

Mikhail S. Storozhevykh[1], and Vladimir A. Yuryev[1]

[1] A. M. Prokhorov General Physics Institute of the Russian Academy of Sciences, 38 Vavilov Street, 119991 Moscow, Russia

[2] Institute of Nanotechnology of Microelectronics of the Russian Academy of Sciences, 32A Leninsky Prospekt, 119991 Moscow, Russia

Corresponding author:

Larisa V. Arapkina arapkina@kapella.gpi.ru

## Abstract

The results of RHEED, FTIR and Raman spectroscopy study of silicon and germanium films with the thickness up to 200 nm grown from molecular beams on dielectric $Si_3N_4$/$SiO_2$/Si(001) substrates are presented. Noticeable changes of the intensity of the N–H and Si–N absorption bands have been observed in the IR absorbance spectra as a result of the deposition of the silicon and germanium films. The thicker was the deposited film, the more considerable were the decrease of N–H absorption band intensity and the increase in that of the Si–N band. This tendency has been observed during the growth of both amorphous and polycrystalline Si or Ge films. The reduction of IR absorption at the band assigned to the N–H bond vibration is explained by breaking of these bonds followed by the diffusion of the hydrogen atoms from the $Si_3N_4$ layer into the growing film of silicon or germanium. The effect of the deposited film thickness on the diffusion of hydrogen is discussed within a model of the diffusion of hydrogen atoms controlled by the difference in chemical potentials of hydrogen atoms in the dielectric $Si_3N_4$ layer and the growing silicon or germanium film. Hydrogen atoms escape from the $Si_3N_4$ layer only during the deposition of a Si or Ge film when its thickness gradually grows. The interruption of the film growth stops the migration of hydrogen atoms into the film because of the decline in the chemical potential difference.

# 1. Introduction

Currently, silicon nitride ($Si_3N_4$) films are extensively used in micro and optoelectronics as insulation and passivation layers and antireflection coatings [1–3]. Low pressure chemical vapor deposition (LPCVD) silicon nitride layers formed at high temperatures (around 700–850°C) have undoubted advantages compared to low temperature (< 450°C) plasma enhanced chemical vapor deposition (PECVD) amorphous ones, which are customarily used in industrial silicon technology. LPCVD $Si_3N_4$ films possess nearly stoichiometric composition and have low hydrogen atom content that ensure their superior electrical properties and thermal stability in comparison with PECVD $Si_3N_4$ ones. For example, LPCVD $Si_3N_4$ layers can be used to protect III–N structures *in situ* to avoid exposure to the atmosphere that would degrade their electrical parameters [4, 5].

A base of passivation process is the diffusion of hydrogen atoms from a silicon nitride film into an interface domain of an active element, e.g., $Si_3N_4/Si$ or $Si_3N_4/SiO_2$ structure, and the inactivation of electrically active centers (traps) there [2, 3, 6]. The understanding of mechanisms of hydrogen atoms diffusion through $Si_3N_4$ layer is important not only for these applications but also for manufacturing of a structure, in which dielectric films are working elements, e.g., in the metal–nitride–oxide–silicon and metal–oxide–nitride–oxide–silicon structures [1, 7]. Hydrogen atoms form bonds with nitrogen or silicon atoms in $Si_3N_4$, which are deep traps for migrating hydrogen atoms. Breaking of the N–H and Si–H bonds and escaping of hydrogen atoms from a $Si_3N_4$ layer could lead to the formation of dangling bonds of N and Si atoms, which become deep traps for charge carriers [7–9].

In this article, we present results the study continuing our previous explorations of diffusion processes taking place during the growth of the silicon and germanium films on dielectric substrates with the topmost layer of LPCVD $Si_3N_4$ [10, 11]. Previously, we have found that the diffusion of hydrogen atoms from the top $Si_3N_4$ layer into the growing Si or Ge films has taken place during the deposition of a silicon or germanium film on a $Si_3N_4/SiO_2/Si(001)$ substrate. This process was observed by Fourier-transform infrared (FTIR) spectroscopy. The simultaneous diffusion of germanium atoms into the $Si_3N_4$ layer was detected using X-ray photoelectron spectroscopy. Both these processes are enhanced by increasing temperature. In this paper, the silicon and germanium films of different thickness have been explored by means of FTIR and Raman spectroscopy, and reflected high-energy electron diffraction (RHEED). It is known that an increase in the thickness of a polycrystalline film might lead to changes in its structure, namely the increase in grain sizes owing to the increase of the film thickness [12]. Polycrystalline silicon and germanium films usually have the columnar structure and their growth on an amorphous surface starts from fine grains, which gradually expand to large sizes [13–16]. The use of samples with different thickness of a deposited Si or Ge film allowed us to explore an influence of the film structure and probable strain in the film on the diffusion of hydrogen atoms from the $Si_3N_4$ layer; we have also put forward a model accounting for the observed phenomenon.

## 2. Sample preparation, experimental methods and equipment

### 2.1. Preparation of samples

Experimental samples were prepared by depositing the Si or Ge layers on $Si_3N_4/SiO_2/Si(001)$ dielectric substrates. A detailed description of the used techniques of the sample production including preliminary chemical treatments of $Si_3N_4/SiO_2/Si(001)$ substrates can be found in our previous work [10]. Now, it is important to notice that the top $Si_3N_4$ layers were formed by CVD using a monosilane-ammonia mixture at the temperature of 750°C and could contain about 8 % of hydrogen atoms bonded with nitrogen atoms [17, 18]. After the chemical treatment, the substrates were loaded in the ultrahigh-vacuum preliminary preparation chamber of the Riber SSC2 center, where they were annealed at 600°C for 6 hours, and then they were transferred into the growth chamber. During the preliminary annealing, the pressure did not exceed $5\times10^{-9}$ Torr in the chamber.

The Si and Ge films were deposited from molecular beams in an ultra-high vacuum EVA 32 (Riber) molecular-beam epitaxy (MBE) chamber using solid sources with electron beam evaporation of Si and Ge. Before the processes, the growth chamber had the pressure of about $4\times10^{-11}$ Torr, which did not exceed $2\times10^{-9}$ Torr during the deposition. Si and Ge deposition rates were about 0.3 Å/c. The film deposition was controlled with the Inficon XTC751-001-G1 (Leybold-Heraeus) film thickness monitor. The thicknesses of the Si layers were 20 Å, 50 and 200 nm and those of Ge films were 20 Å, 50, 100 and 200 nm. The samples with the thin (50 and 100 nm) and thick (200 nm) films were subjected to the heat treatment for the same total time due to the additional annealing of the thinner samples after the flux shutoff of Si or Ge atoms. The film growth temperatures were 30, 450 and 500°C. Samples were heated with tantalum radiative heaters from the rear side both in the pretreatment chamber and in the MBE chamber. The growth temperature was controlled using chromel-alumel and tungsten-rhenium thermocouples of the heaters. An IMPAC IS12-Si infrared pyrometer (LumaSense Technologies) was used for temperature monitoring and thermocouple graduation. The growth surface structure was monitored using an RH20 RHEED tool (Staib Instruments). An RGA-200 residual gas analyzer (Stanford Research Systems) was used for residual atmosphere monitoring before and during the deposition process.

## 2.2. Analytical techniques

The main experimental methods were RHEED, FTIR spectral analysis and Raman spectroscopy.

The RHEED images were collected *in situ* during the deposition processes and after sample cooling to the room temperature. The RHEED patterns were processed as those of polycrystalline silicon and germanium surfaces.

FTIR transmission and reflection spectra of the samples were explored using a vacuum IFS-66v/S spectrometer (Bruker) with the spectral resolution of 10 cm$^{-1}$. The spectrometer was pumped out to the pressure of 2 mbar during spectra recording that enabled a considerable reduction of the interferences from carbon dioxide and water vapor bands. A detailed description of the measurement technique can be found in our previous works [10, 11]. IR absorbance spectra were analyzed for peaks using Gaussian functions.

Raman spectra were registered was performed by using an inVia Qontor spectrometer (Renishaw). The spectra acquisition was realized using vertically polarized laser excitation at the wavelengths of 405, 532, 633 and 785 nm with the emission power of 15, 0.1, 1.7 and 5.4 mW, respectively. A diffraction grating with 1200 gr/mm was used for 785-nm and 633-nm lasers and that with 2400 gr/mm was applied for 405-nm and 532-nm ones; a 50× objective was employed in all the experiments. Using lasers with several wavelengths enabled obtaining the information from different depths in the sample. The measurements were carried out at room temperature. Voigt functions were used for the peak analysis of Raman bands.

## 3. Results

### 3.1. RHEED

RHEED technique was used to characterize the surface of growing crystalline silicon and germanium films. Figure 1 and Figure 2 present RHEED patterns of the surface of the 20 Å, 50 nm and 200 nm thick Si and Ge films grown at 500°C, respectively. All the RHEED patterns consist of diffuse Debye rings that confirm that the formed Si and Ge films are polycrystalline [19]. The radii of the rings correspond to the reciprocal interplanar distances ($1/d_{hkl}$) in the lattices of silicon or germanium [20], which have the space group $Fd\bar{3}m$ (227). In Figure 1a and Figure 2a, the RHEED images are related to the thinnest samples comprising 20 Å thick films of Si and Ge, respectively. All rings are seen to be continuous, which means that crystalline grains forming the films randomly oriented. In the pattern of the Si film (Figure 1a), broad intense reflexes are observed, so that neighboring rings overlap each other and cannot be resolved (see, e.g., the (422) and (333) pair). For the Ge film (Figure 2a), the rings are more intense and narrower than for the Si layer and adjacent reflexes are quite resolvable. The comparison of the 20 Å thick Si and Ge samples shows that the Ge films consisted of larger grains at the initial growth stage than the Si layers did. With an increase in the thickness of the Si film (Figure 1b and c), the rings become narrower and discontinuous ones appear that indicate the formation of coarser grains with a preferential growth orientation.

Another scenario is observed for the Ge deposition. In this case, broken rings form with the increasing Ge film thickness as well (Figure 2b and c), but a strong background appears which hinders resolving rings. Possibly, this is connected with strong scattering of electrons on the surface relief. The 50, 100 and 200 nm thick germanium films had the matte surface, whereas all silicon ones always were glossy.

Thus, as the film thickness increased, the coarsening of the grains forming the polycrystalline silicon or germanium layers occurred.

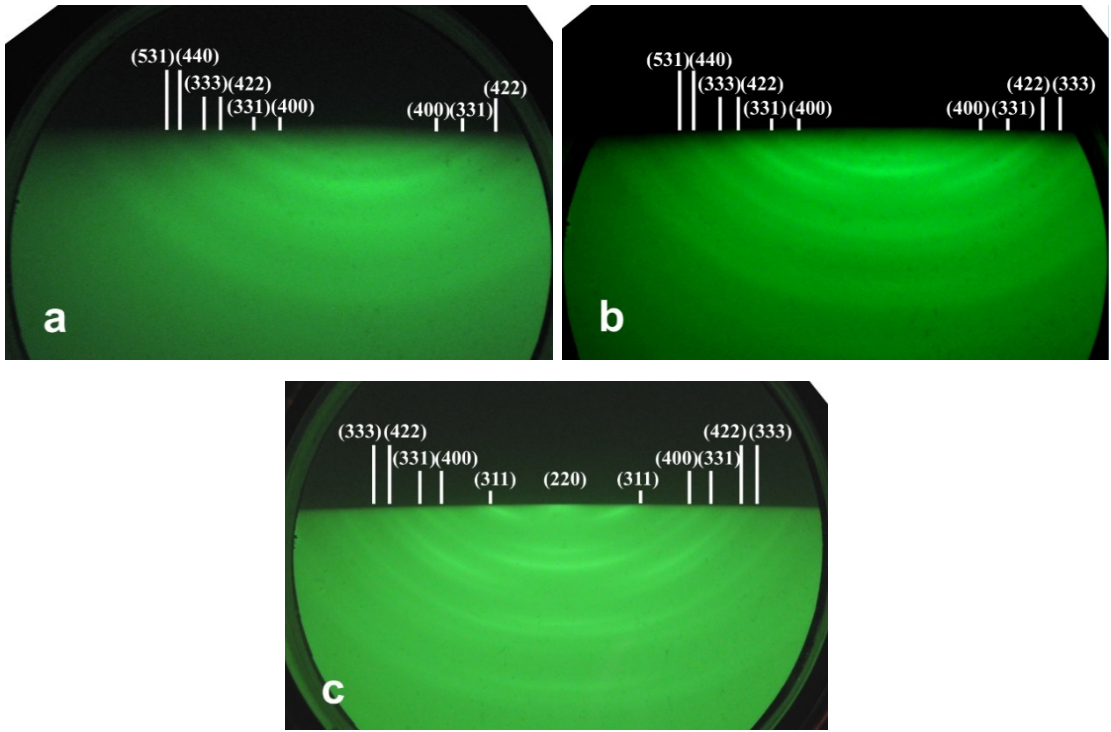

Figure 1. RHEED patterns of the surface of a Si layer grown at 500°C; the film thicknesses are (a) 2, (b) 50 and (c) 200 nm.

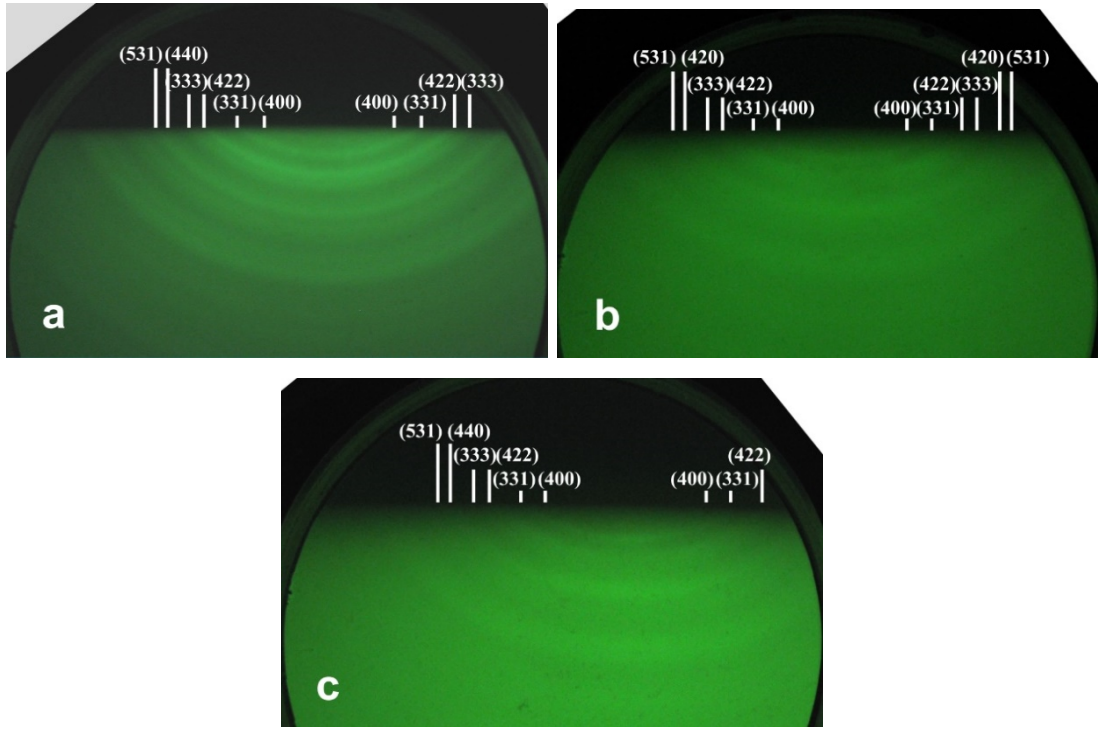

Figure 2. RHEED patterns of the surface of a Ge layer grown at 500°C; the film thicknesses are (a) 2, (b) 50 and (c) 200 nm.

## 3.2. Raman spectroscopy

### 3.2.1. Si layers

For the Raman spectroscopy, the samples with the 50 and 200 nm thick polycrystalline silicon (poly-Si) films were used. The Raman spectra obtained using different lasers for the sample with the 200 nm thick poly-Si layer grown at 450°C are depicted in Figure 3. The feature of this sample was that in the middle of its growth (at 100 nm thickness), the transition from an amorphous to a polycrystalline structure was observed. The high-resolution TEM image of this sample was presented in our previous article [21]. The Raman spectra exited with the 532-nm, 633-nm and 785-nm lasers contain the vibrational bands around 480 cm$^{-1}$ and 520 cm$^{-1}$ assigned to the TO phonon of amorphous silicon (α-Si) and the TO phonon of crystalline silicon (c-Si), respectively [22, 23]. The radiation of the 785-nm laser penetrated deep into the substrate, and the corresponding spectrum contains an additional 2TA phonon peak, typical of crystalline silicon [23, 24]. The radiation of the 405-nm laser did not pass through the ~100 nm thick polycrystalline domain of the sample, and the spectrum contains only one band centered at ~ 520 cm$^{-1}$, assigned to TO(c-Si).

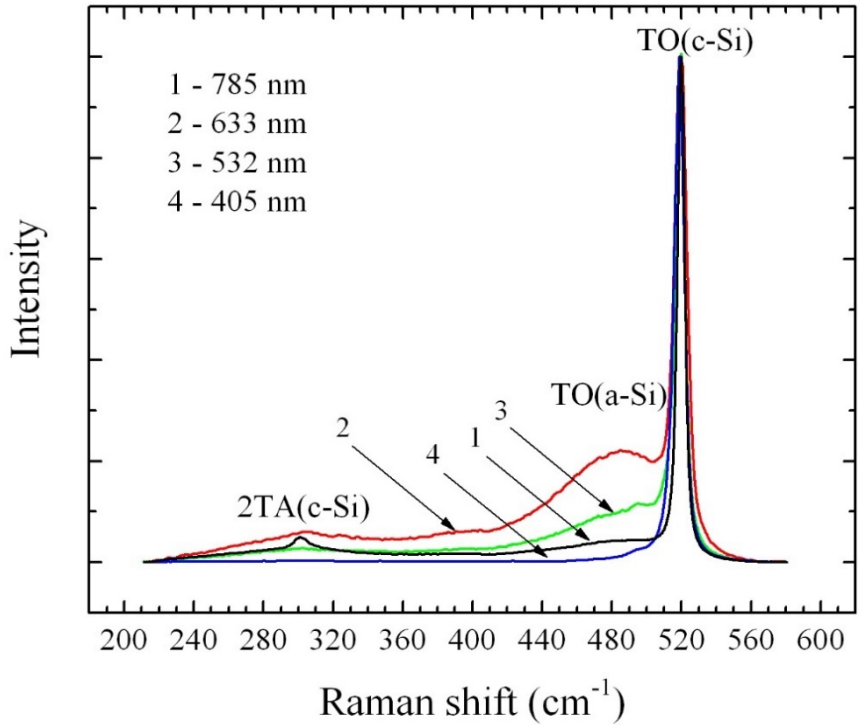

Figure 3. Comparison of the Raman spectra of the 200 nm thick Si layer grown at 450°C for different wavelength of the exciting laser; the spectra are normalized to the maximum of the strongest band peaked at 520 cm$^{-1}$.

Figure 4 a demonstrates a result of the deconvolution of the band at ~520 cm$^{-1}$ related to the polycrystalline part of the grown film (the 405-nm laser was used to excite Raman scattering). Strong vibrational bands peaked at ~519 cm$^{-1}$ (Peak 1), ~518 cm$^{-1}$ (Peak 2) and a broad weak band peaked around 500 cm$^{-1}$ (Peak 3) have been obtained. The position and FWHM of Peak 1 are similar to those of the TO(c-Si) vibration band obtained for the $Si_3N_4/SiO_2/Si(001)$ substrate and marked as c-Si in Figure 4. Peak 2 and Peak 3 may be assigned to vibrational bands of the poly-Si layer and connected with grain boundaries [25–31].

The peak analysis of Raman spectra obtained using 532-nm, 633-nm and 785-nm lasers has revealed the same peaks as those shown in Figure 4a and one additional wide peak around 480 cm$^{-1}$ (Peak 4) corresponding to the TO(α-Si) vibration (Figure 4b). The area of Peak 3 in the spectra increased (Figure 4) with a change in the laser wavelength from 405 to 532 nm, when the laser radiation approached the interface with the amorphous part of the film and to the region with a larger area of grain boundaries. The Raman spectra excited with the 532-nm and 633-nm lasers showed practically the same results of the peak fitting with the exception of the intensity of the peak at ~480 cm$^{-1}$, which was stronger in the spectrum excited with the 633-nm laser. In the Raman spectrum obtained with the 785-nm laser, a strong influence of light scattering by the Si–Si bonds due to the penetration of the laser radiation to the substrate was observed. With this excitation, the strongest TO(c-Si) peaks and weak TO(α-Si) ones were observed. Thus, we can suppose that Peak 1 relates to the TO(c-Si) vibration band of the crystalline silicon phase in the poly-Si film (i.e. to interior parts of grains), whilst Peak 2 and Peak 3 appear due to Raman scattering at Si grain boundaries.

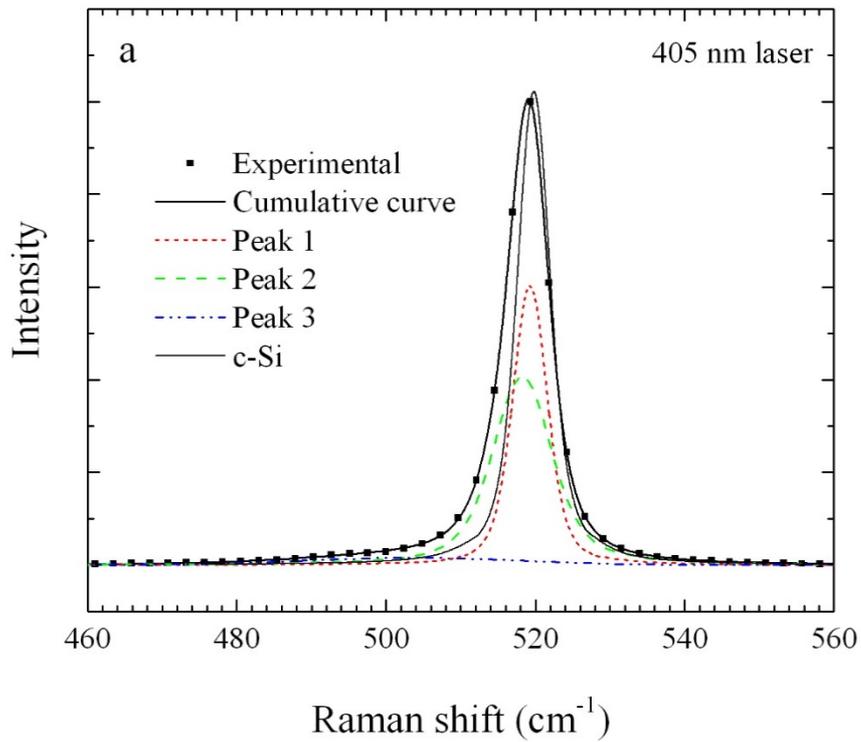

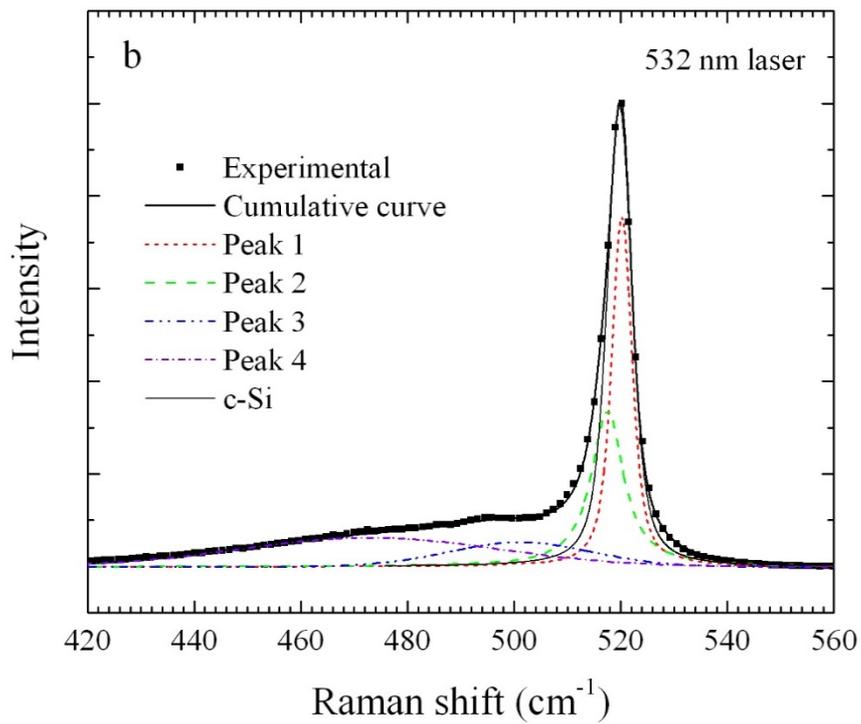

Figure 4. Deconvolution of Raman spectra of the 200 nm thick Si layer grown at 450°C in the vicinity of the TO(c-Si) vibration band peaked at 520 cm$^{-1}$; the c-Si spectrum relates to the $Si_3N_4/SiO_2/Si(001)$ substrate; the laser wavelengths

are 405 nm (a) and 532 nm (b); the spectra are normalized to the maximum of the strongest peak at 520 cm$^{-1}$.

The analysis of the Raman spectra of the sample with 200 nm thick Si film grown at 500°C (Figure 5), which is polycrystalline across the entire thickness, have shown that only the radiation of the 785 nm laser definitely penetrated the substrate that led to the appearance of the 2TA(c-Si) peak at ~300 cm$^{-1}$ in the spectrum. It is safe to state that the radiation of the 532-nm and-633 nm lasers passed through the 200 nm thick poly-Si layer, although the 2TA(c-Si) peak is faintly visible in the spectra of Raman scattering they excite. The radiation of 405-nm laser was completely absorbed by the 200 nm thick poly-Si layer, just like by the 100 nm thick one. All spectra of the 200 nm thick poly-Si layer contain a strong TO(c-Si) vibration band at ~520 cm$^{-1}$. The deconvolution of this band regardless the used laser has revealed three peak components (Figure 6), which are similar to those shown in Figure 4a. With the increase of the laser wavelength from 405 nm to 633 nm, the intensity of Peak 3 notably increased (Figure 6b) that was connected with the increasing influence of domain having greater area of grain boundaries. In the Raman spectrum for the 785-nm laser, the intensity of the TO(c-Si) vibration band (Peak 1) significantly increased (Figure 5) that reduced the number of registered peaks. The positions of Peak 1, Peak 2 and Peak 3 did not depend on the exciting laser wavelength.

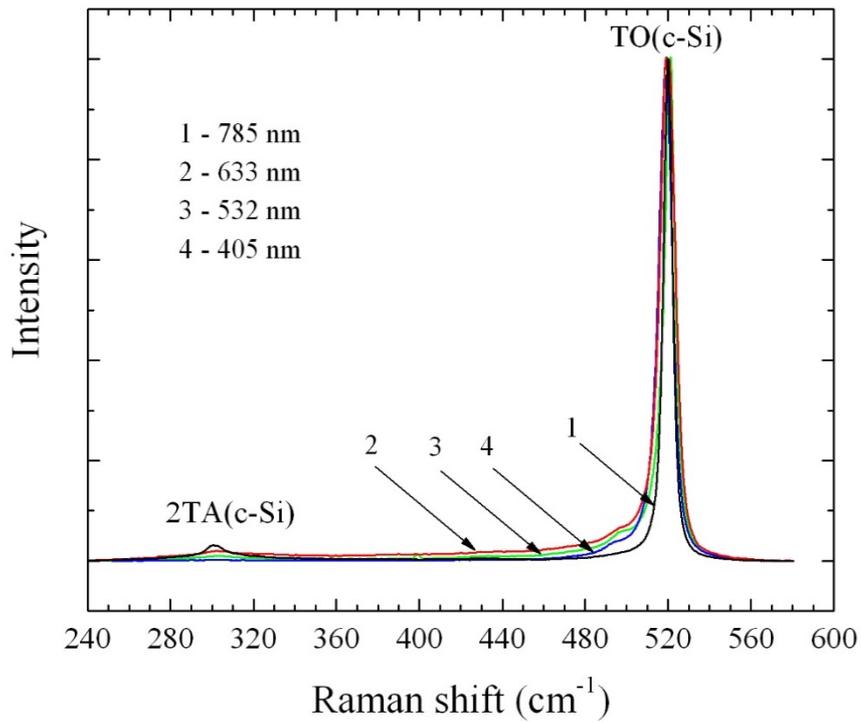

Figure 5. Comparison of the normalized spectra of Raman scattering excited with 405-nm, 532-nm, 633-nm and 785-nm lasers in the sample with 200 nm thick Si layer grown at 500°C; the spectra are normalized to the maximum of the strongest peak at 520 cm$^{-1}$.

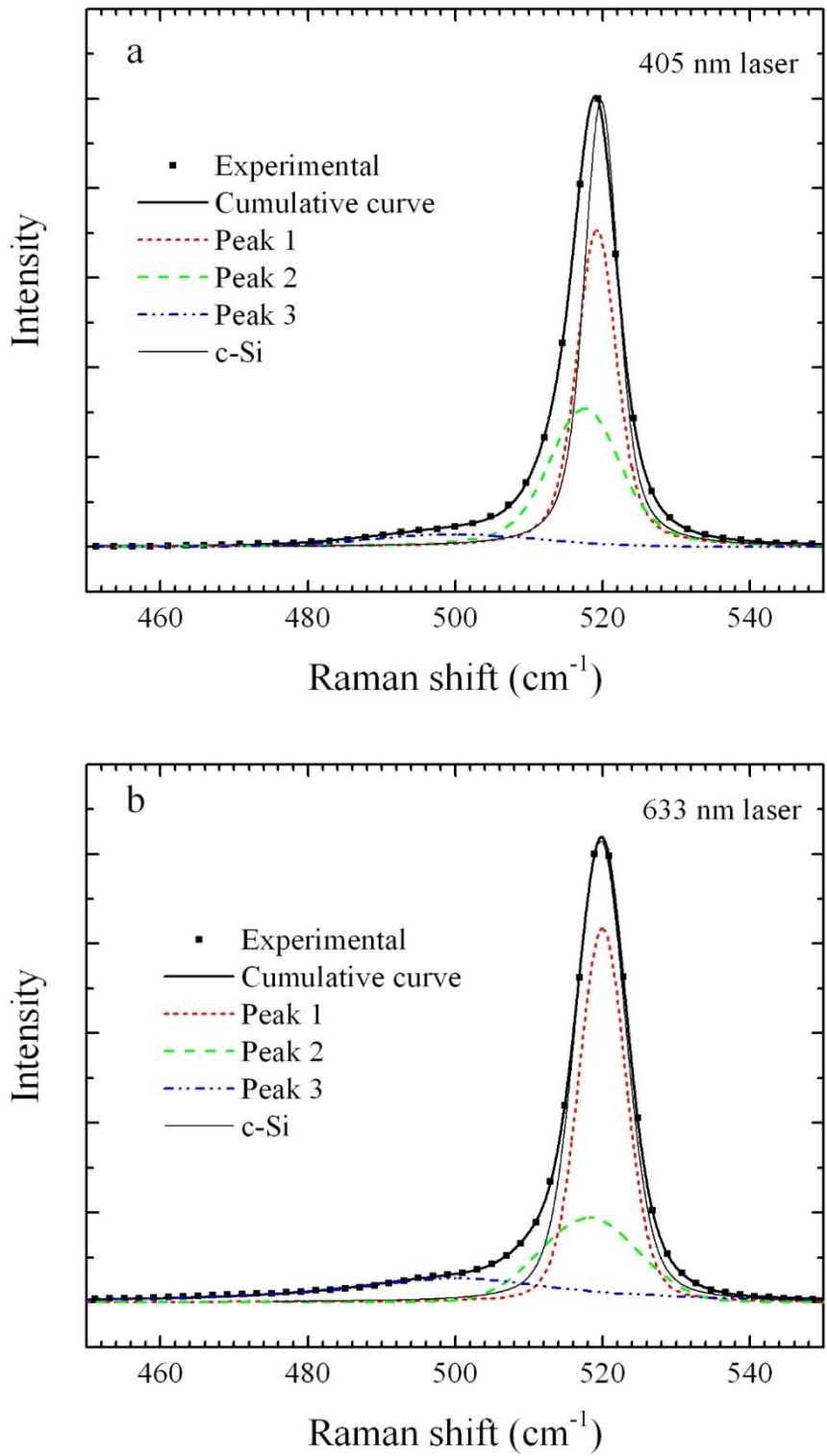

Figure 6. Peak analysis of Raman spectra nearby TO(c-Si) vibration band of 200 nm thick Si layer grown at 500°C; the c-Si spectrum is associated with the $Si_3N_4/SiO_2/Si(001)$ substrate; laser wavelengths are 405 (a) and 633 nm (b); the spectra are normalized to the maximum of the strongest peak at 520 cm$^{-1}$.

Analysis of the Raman spectra of the 50 nm thick poly-Si film grown at 500°C has shown that the radiation of the 532-nm, 633-nm and 785-nm lasers penetrated the substrate and excited the 2TA(c-Si) peak. In the Raman spectrum excited with the 405-nm laser the 2TA(c-Si) peak was absent. Besides, it was also absent in the spectra of the $Si_3N_4/SiO_2/Si(001)$ and $Si(001)$ substrates, therefore it is impossible to say definitely about the complete absorption of radiation of the 405-nm laser by the 50 nm poly-Si layer. For all the spectra, the main band peaked around 520 $cm^{-1}$ was decomposed to several peaks. There were three components (Peak 1, Peak 2 and Peak 3) in the spectra registered using the 405-nm and 532 nm lasers, two peaks (Peak 1 and Peak 3) in the spectrum recorded with 633-nm laser and a single peak (Peak 1) in the spectrum obtained with the 785-nm laser, with the positions of these peaks been the same as those in the spectra of the 200 nm thick poly-Si layer (Figure 4 and Figure 6). The reducing of the number of peaks in the spectra is accounted for the increasing influence of light scattering in the $Si_3N_4/SiO_2/Si(001)$ substrate. The comparison of the Raman spectra of the 50 nm and 200 nm thick poly-Si layers excited with the 405-nm laser are presented in Figure 7. It is seen that the spectrum of the 50 nm thick sample has a higher intensity of emission around 500 $cm^{-1}$ (Peak 3) that agrees with our RHEED results, according to which the film domain near the interface with the substrate consists of finer grains compared to the near-surface area of the 200 nm thick film.

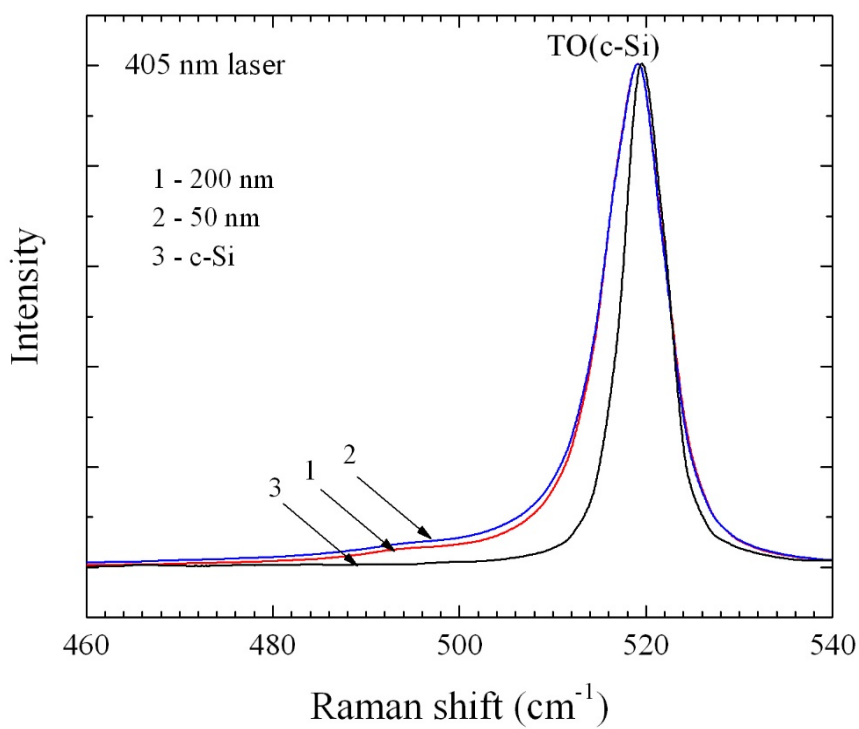

Figure 7. Comparison of the normalized Raman spectra of samples with 50 nm and 200 nm thick Si layers grown at 500°C. Laser wavelength is 405 nm. The c-Si spectrum belongs to the $Si_3N_4/SiO_2/Si(001)$ substrate. The spectra are normalized to the maximum of the highest peak at 520 cm$^{-1}$.

**3.2.2. Ge layers**

The samples with the 50 nm, 100 nm and 200 nm thick Ge films grown at 500°C on the $Si_3N_4/SiO_2/Si(001)$ substrate were used for the Raman spectroscopy study. According to RHEED results, all these films were polycrystalline (Figure 2).

Comparison Raman spectra obtained using three different lasers with the wavelengths of 532 nm, 633 nm and 785 nm has shown that the exciting radiation passed though the samples with 50 nm and 100 nm thick poly-Ge films and was scattered in the substrate that led to the appearance of the TO(c-Si) peak in the spectra. Regarding 200 nm thick sample, only the radiation of the 785-nm laser penetrated into the substrate. In all the spectra, the strongest vibrational band was that peaked at 300 cm$^{-1}$.

A result of the deconvolution of the vibrational band at ~300 cm$^{-1}$ excited with the 785 nm laser for the 200 nm thick the poly-Ge sample is shown in Figure 8. The band is composed of two components: a strong narrow peak at ~300 cm$^{-1}$ (Peak 1) and a wide weak band near 290 cm$^{-1}$ (Peak 2). Within the measurement error, the position and FWHM of Peak 1 coincide with those of the TO(c-Ge) in the crystalline germanium (c-Ge) [32, 33]. The deconvolution of the Raman spectra obtained using of the 532-nm (Figure 8b) and 633-nm lasers revealed the same peak components as those presented in Figure 8a. The value of the ratio of peak areas $S_{Peak\ 1}/S_{Peak\ 2}$ decreased at the transition from 532-nm to 785-nm laser. With increasing depth of penetration, the laser radiation reached regions containing finer grains, a large area of boundaries of which enhanced their contribution to Raman scattering. This allows us to assume that the vibrational band at ~290 cm$^{-1}$ (Peak 2) could be connected with grain boundaries in poly-Ge [34].

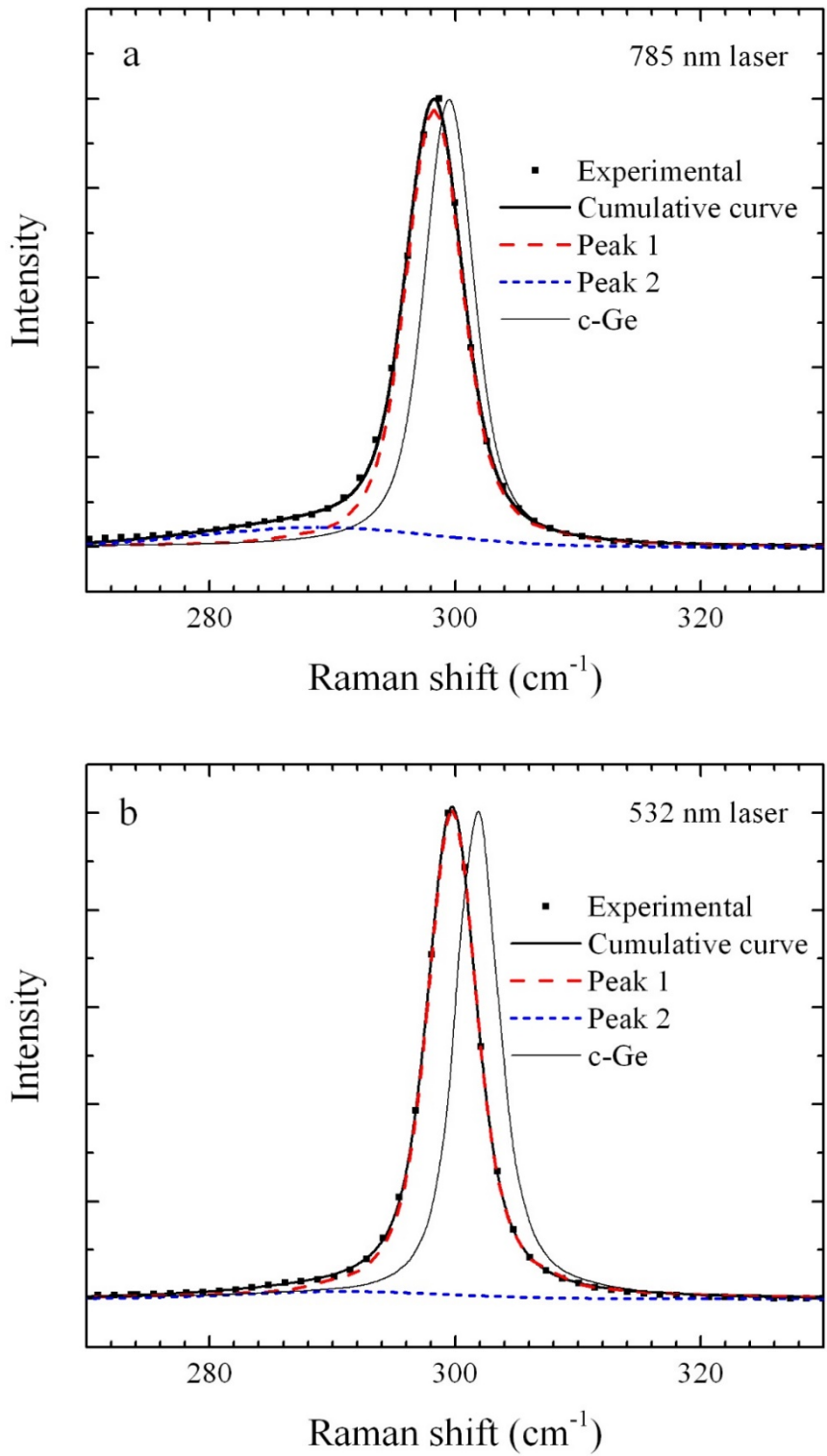

Figure 8. Result of the deconvolution of Raman spectra around the peak at 300 cm$^{-1}$ of the sample with the 200 nm thick poly-Ge film grown at 500°C; the c-Ge spectrum is related to crystalline germanium; laser wavelengths are 785 nm (a) and 532 nm (b); the spectra are normalized to the maximum of the strongest peak at 300 cm$^{-1}$.

Figure 9a shows a result of comparison of the Raman spectra of the samples with the 50, 100 and 200 nm thick poly-Ge films obtained with the 785-nm laser. All the spectra have one strong band at ~300 cm$^{-1}$, the position of which does not depend on the sample thickness. For the thinner samples, the increase in the intensity of the spectra near 290 cm$^{-1}$ is observed. The analysis of the vibrational band at ~300 cm$^{-1}$ of the Raman spectra of the 50 and 100 nm thick poly-Ge films revealed that regardless of the used exciting laser it consisted of the same peak components as ones shown in Figure 8. The result of the deconvolution of the spectrum of the 50 nm thick sample is presented in Figure 9 b: there are two peaks composing the band (Peak 1 and Peak 2), with the position and FWHM of the Peak 1 being closer to values for the c-Ge. With a decrease in the film thickness from 200 to 50 nm, a reduction of the peak area ratio $S_{Peak\ 1}/S_{Peak\ 2}$ has been observed (Figure 8 a and Figure 9 b) that we connect with an increase in content of fine grains with a decrease in film thickness. Remind that the use of the 633-nm instead of 532-nm laser resulted in the similar change of the ratio of peak areas. Consequently, thinner films are formed by finer grains, and their boundaries make a greater contribution to the Raman spectra. This also supports our assumption that the vibrational band at ~290 cm$^{-1}$ (Peak 2 in Figure 8 and Figure 9) is due to Raman scattering on grain boundaries in the poly-Ge films.

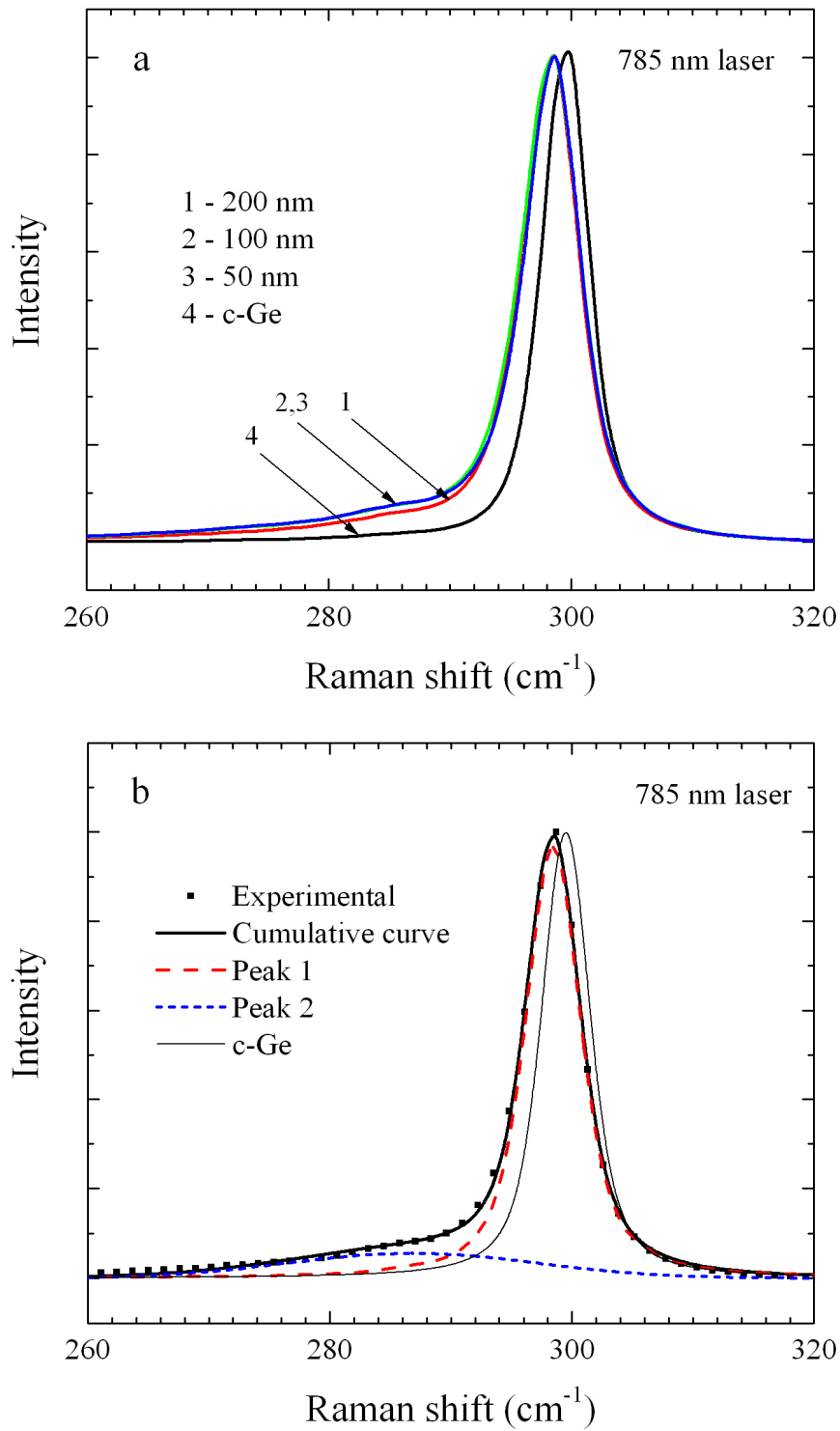

Figure 9. Raman spectra nearby the 300 cm$^{-1}$ peak of the samples with different film thickness grown at 500°C (a) and a result of the deconvolution of the Raman spectra of the sample with the 50 nm thick poly-Ge film (b).

Concerning the α-Ge films grown at 30°C with the thickness of 50 and 200 nm, their Raman spectra, as we have observed, completely match each other and quite resemble ones usually registered at amorphous germanium layers [11, 36].

The Raman observations enable supposing that the analyzed polycrystalline silicon and germanium films having different thickness below 200 nm are characterized by the same values of internal stresses emerged in them. The same result was obtained for the amorphous germanium films with different thickness. The polycrystalline film samples have the oriented structure formed by grains with crystalline interior, with the sizes of grains increasing with increasing layer thickness.

### 3.2. FTIR spectroscopy

IR absorbance spectra of the samples with Ge and Si films were recorded in the range from 400 to 4600 cm$^{-1}$. Figure 10 presents IR absorbance spectra for the samples with the 50, 100 and 200 nm thick poly-Ge films (grown at 500°C) in the range from 650 to 1350 cm$^{-1}$. For comparison, the spectrum recorded at $Si_3N_4/SiO_2/Si(001)$ substrate is also shown. The spectra are seen to comprise two wide bands peaked at ~825 and ~1090 cm$^{-1}$. Both of them are assigned to the Si–O bond of vibrations originating in the $SiO_2$ layer of the $Si_3N_4/SiO_2/Si(001)$ substrate [35]. Each of them has a broad shoulder spreading up to nearly 1000 cm$^{-1}$ for the band at ~825 cm$^{-1}$ and 1300 cm$^{-1}$ for that at ~1090 cm$^{-1}$. IR absorbance spectra of the samples with the 50 and 200 nm thick poly-Si films grown at 500°C and with the 50 and 200 nm thick α-Ge films grown at 30°C had the similar shape. The same IR spectra recorded for Si and Ge films grown at different temperatures could be found in our previous works [10, 11].

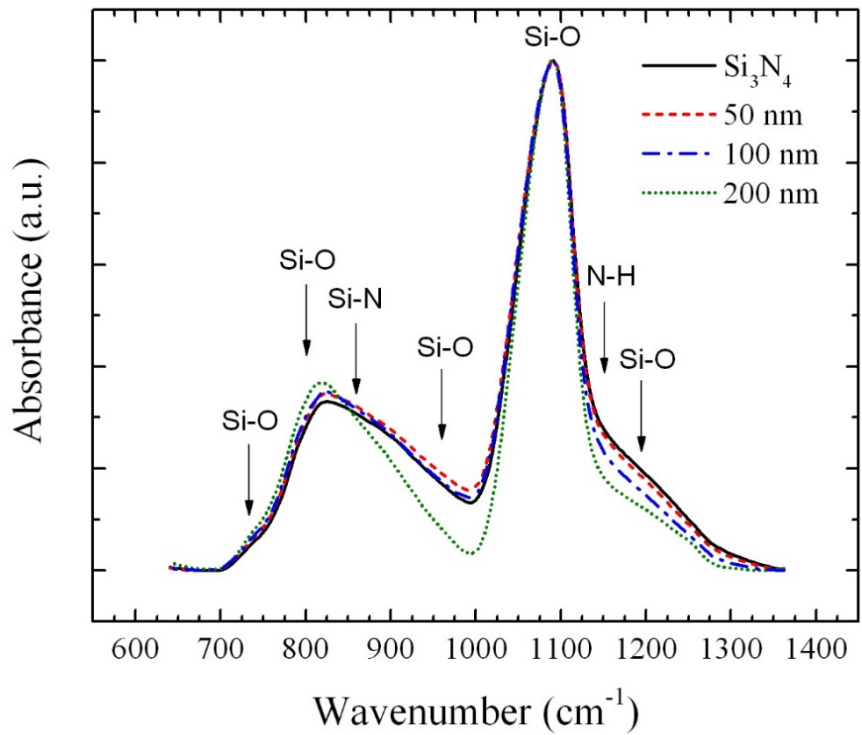

Figure 10. FTIR spectra of the poly-Ge films grown at 500°C on the $Si_3N_4/SiO_2/Si(001)$ substrate; the film thicknesses are 50, 100 and 200 nm; the spectra are normalized to the maximum of the peak at 1090 cm$^{-1}$; the solid line relates to the $Si_3N_4/SiO_2/Si(001)$ substrate.

Figure 11 depicts the result of spectrum decomposition using eight Gaussian peaks for the sample with the 100 nm thick poly-Ge film. All IR absorbance spectra have been processed as well. The IR spectra consist of three types of bond vibrations. The lines peaked at 735, 805, 960, 1075, 1100 and 1190 cm$^{-1}$ are the vibrational bands of the Si–O bonds [37, 38]. It should be noted that the peak at ~960 cm$^{-1}$ relates to the vibrational band of the Si–O bonds in the oxynitride layer that covered the surface of the $Si_3N_4/SiO_2/Si(001)$ substrate [11]. The other Si–O vibration bands are assigned to the bonds in the $SiO_2$ layer of the substrate. The peak at ~1150 cm$^{-1}$ is connected with the vibrations of the N–H bonds and one at ~860 cm$^{-1}$ is generated by the vibrations of the Si–N bonds [39, 40]. These peaks are related to vibrations taking place in the $Si_3N_4$ layer of the substrate.

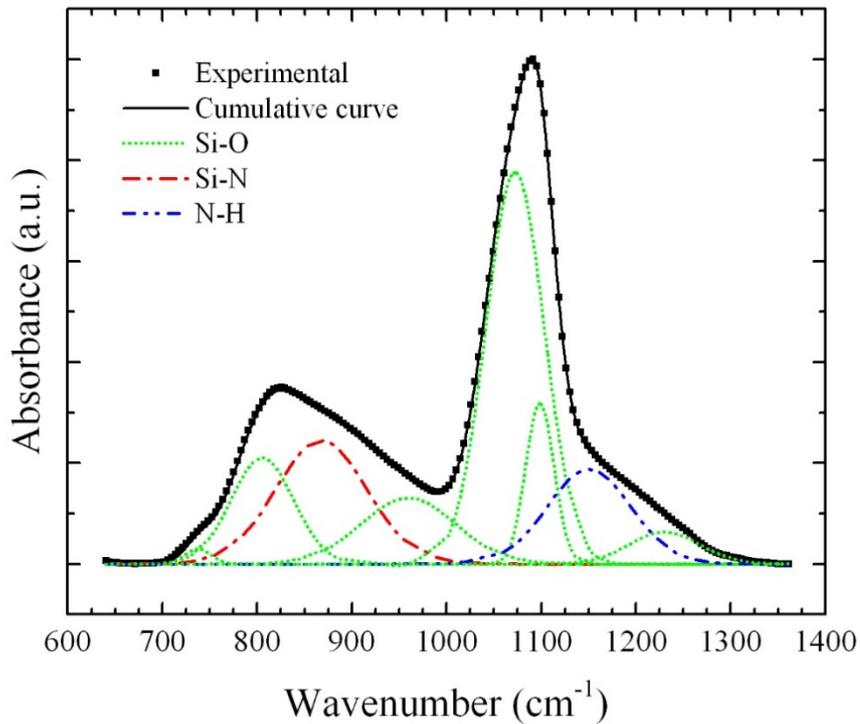

Figure 11. Peak analysis of the FTIR spectrum (650 to 1350 cm$^{-1}$) for the poly-Ge/Si$_3$N$_4$/SiO$_2$/Si(001) sample with 100 nm thick Ge film; Gaussian peaks are assigned to vibrations of the Si–O, Si–N and N–H bonds; the spectrum is normalized to the maximum intensity of the peak at 1090 cm$^{-1}$.

A comparison of the spectra has revealed that the absorption bands relating to the vibrations of the N–H (~1150 cm$^{-1}$), Si–N (~860 cm$^{-1}$) and Si–O (~960 cm$^{-1}$) bonds have changed after the germanium and silicon film deposition. Figure 12 presents results of the IR spectra deconvolution in the form of evolution of the bands assigned to the vibrations of the Si–O, Si–N and N–H bonds. In Figure 12a, the changes in the intensities of the absorption bands for the samples with poly-Ge films grown at 500°C are depicted. It is seen, that with an increase of the film thickness, the strength of the vibrational bands of Si–O and N–H decreases but that of the Si–N vibration band increases. The similar changes are observed for the samples with the poly-Si films grown at 500°C (Figure 12b) and the α-Ge films grown at 30°C (Figure 12c). Comparison of intensity of the N–H absorption bands for the α-Ge and poly-Ge films shows that it drops more significantly in the thin amorphous film grown at 30°C than in the thin polycrystalline film grown at 500°C (Figure 12 a and c). It should be noted that a decrease in the intensity of the Si–O vibration band with an increase in film thickness is more considerable for the samples with the Ge films than for the samples with the Si layers.

Figure 13 presents IR absorbance spectra recorded in the range of 400 to 650 cm$^{-1}$ for the samples containing poly-Ge films of different thickness. The spectra have two broad bands around 460 and 615 cm$^{-1}$. The first of them corresponds to the Si–O bond vibrations and the second is connected with the Si–Si ones. It is seen that with an increase of the grown Ge film thickness, the intensity of the spectra decreases in range from 490 to 580 cm$^{-1}$.

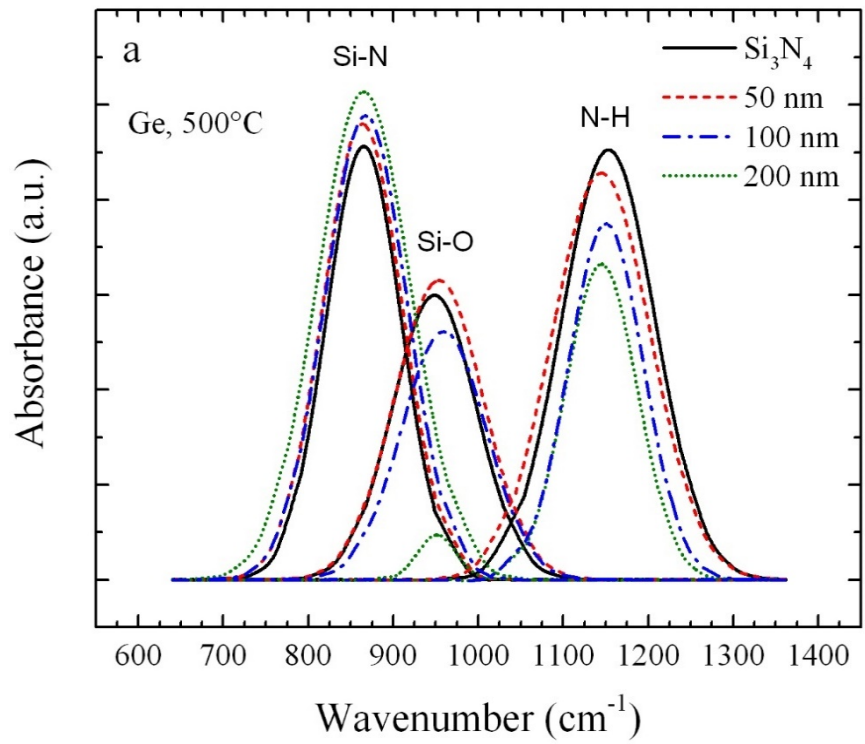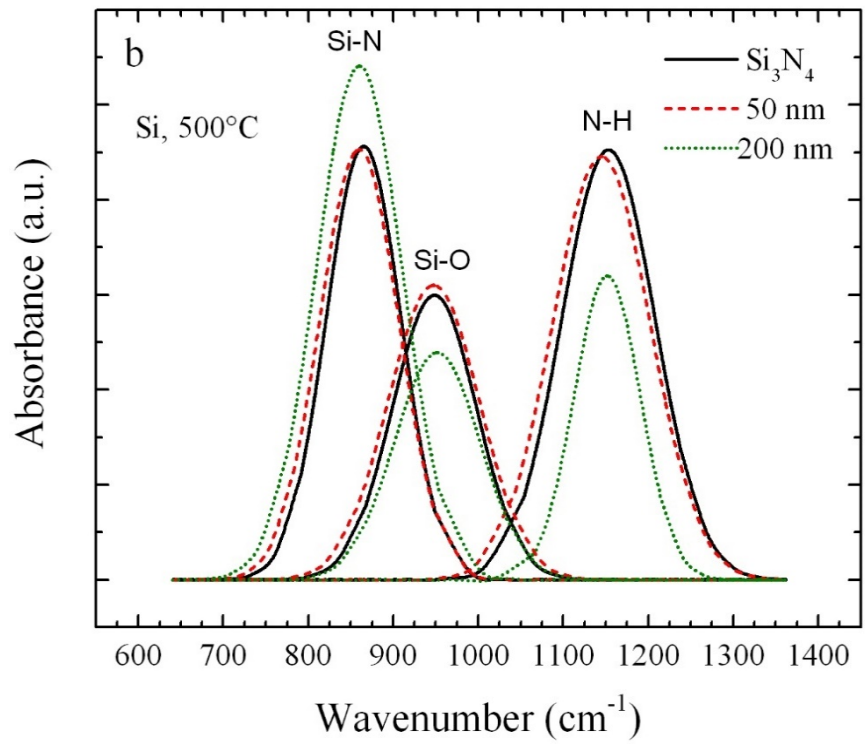

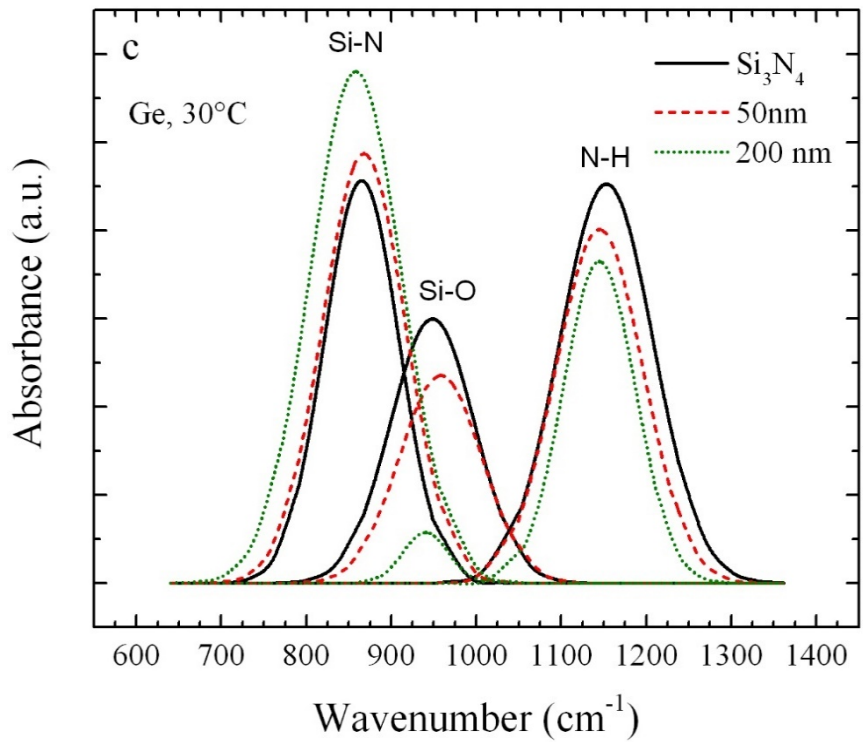

Figure 12. Evolution of the bands related to the vibrations of the Si–O, Si–N and N–H bonds in the FTIR spectra of the samples with different film thickness: (a) poly-Ge films grown at 500°C; (b) poly-Si films grown at 500°C; (c) α-Ge films grown at 30°C; the solid line relates to the $Si_3N_4/SiO_2/Si(001)$ substrate.

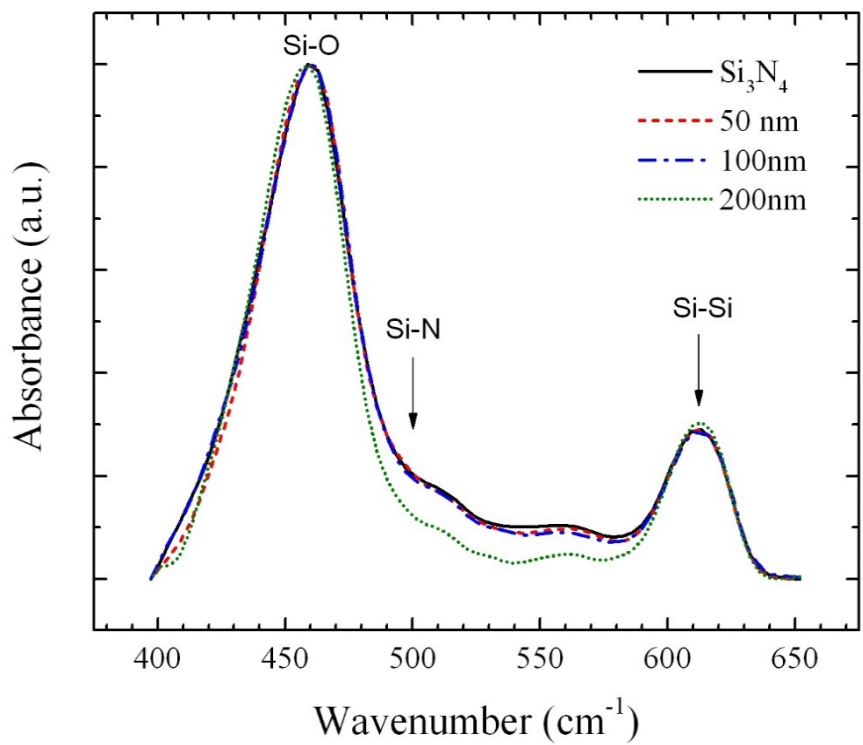

Figure 13. FTIR spectra for the samples with the poly-Ge films grown at 500°C on the $Si_3N_4/SiO_2/Si(001)$ substrate; the film thicknesses are 50, 100 and 200 nm; the spectra are normalized to the maximum intensity of the peak at ~460 cm$^{-1}$; the solid line relates to the $Si_3N_4/SiO_2/Si(001)$ substrate.

Figure 14 presents a result of the deconvolution of the absorbance spectra depicted in Figure 13 for the 100 nm thick poly-Ge film grown at 500°C. The spectrum is composed of six Gaussian peaks. The peaks arranged about 425 and 460 cm$^{-1}$ correspond to the Si–O absorption bands; the peak with the maximum at ~500 cm$^{-1}$ relates to the vibrations of the Si–N bonds [41]; the band at about 615 cm$^{-1}$ is originated from the vibrations of the Si–Si bonds and the peak at ~560 cm$^{-1}$ is unrecognized. All these bands relate to the absorption in the $Si_3N_4/SiO_2/Si(001)$ substrate and the changes in their intensities result from by processes of the deposition of the germanium films. Analogous results were obtained in our previous investigation of silicon films growth on similar substrates [10].

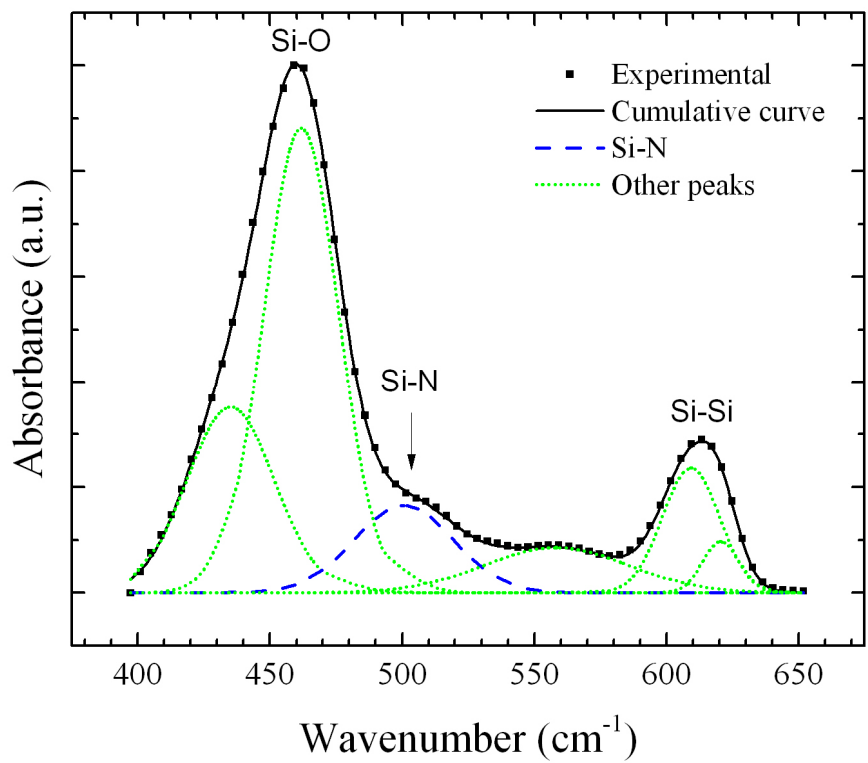

Figure 14. Peak analysis of the FTIR spectrum (400 to 650 cm$^{-1}$) for the samples with poly-Ge films grown at 500°C on the Si$_3$N$_4$/SiO$_2$/Si(001) substrate (Figure 13); the peaks are associated with the vibrations of the Si–O, Si–N and Si–Si bonds; the spectrum is normalized to the maximum intensity of the peak at ~460 cm$^{-1}$.

The changes in the intensity of the Si–N absorption band at ~500 cm$^{-1}$ depending on the thickness of the poly-Ge or α-Ge films are shown in Figure 15. It is seen that with the increase of the grown film thickness, the intensity of the Si–N absorption band decreases and the band gets narrower (Figure 15). The activity of the Si–N absorption band can be connected with an order in the arrangement of atoms in Si$_3$N$_4$ layer. The better the order in the atom arrangement is in the Si$_3$N$_4$ layers, the lower the intensity of this band is [41]. We have observed the similar results in the case of the Si film deposition at different growth temperatures [10]. The growth of the α-Ge films at 30°C leads to the decrease in the Si–N absorption band intensity (Figure 15b); however, the difference between the values obtained for the 50 and 200 nm thick α-Ge films is much less than it is in the case of the poly-Ge films growth (Figure 15a). A gap between the absorption intensity values in the uncoated Si$_3$N$_4$/SiO$_2$/Si(001) substrate and in those coated with Ge is less in the case of the thick α-Ge film but greater in the case of the thin one than it is in the case of the thick or thin poly-Ge film (Figure 15). A similar behavior of the intensity of the Si–N vibration band was observed if Si films were deposited.

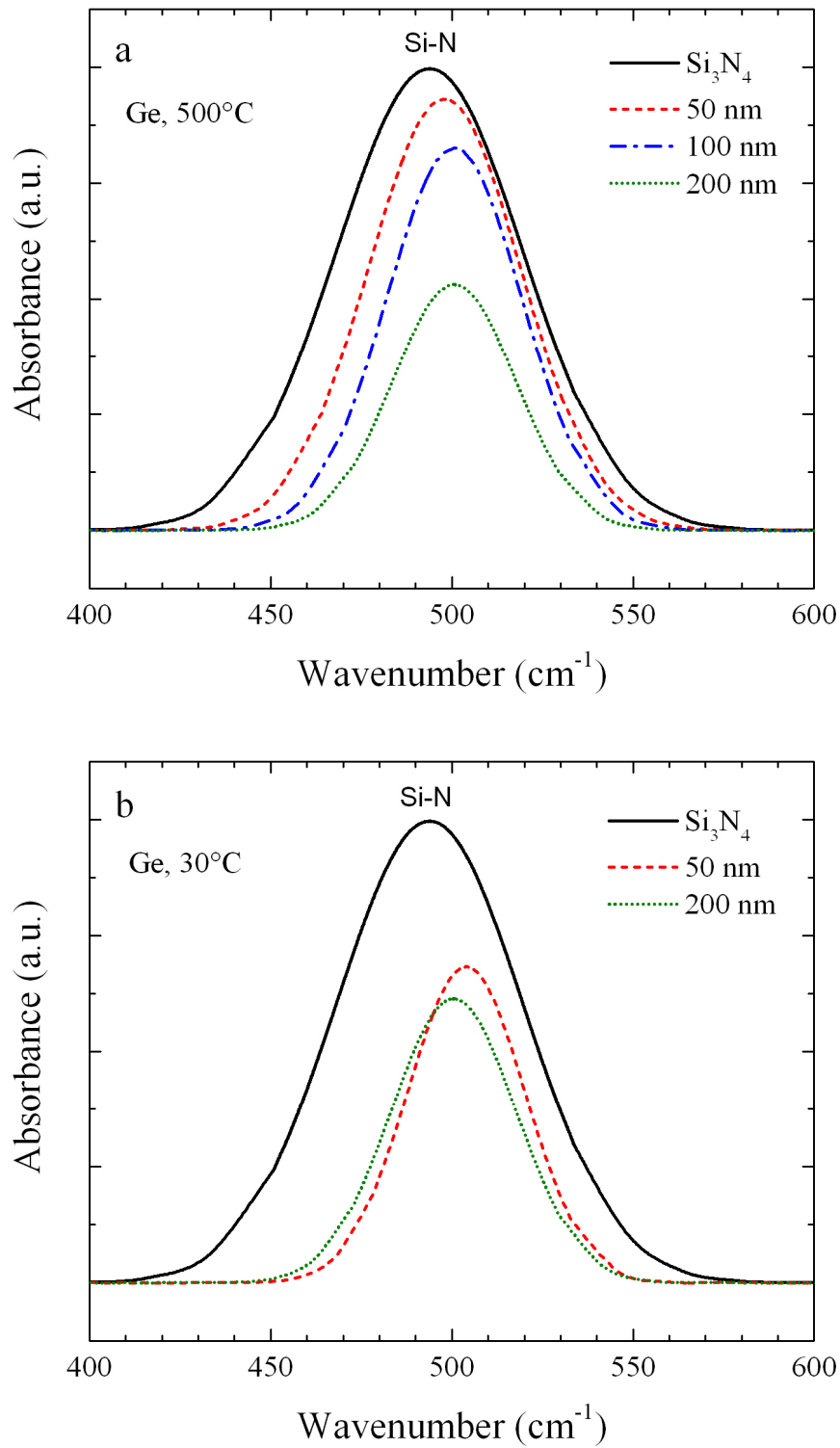

Figure 15. Evolution of the bands assigned to the vibrations of the Si–N bond in the IR spectra for the samples with different thickness: (a) poly-Ge films grown at 500°C; (b) α-Ge films grown at 30°C; the solid line relates to the $Si_3N_4/SiO_2/Si(001)$ substrate.

Figure 16 shows sums of the IR transmittance and reflectance spectra recorded in the wavenumber interval from 1900 to 4600 cm$^{-1}$ for the samples with poly-Ge layers deposited at 500°C. The band observed at ~3340 cm$^{-1}$ relates to the vibrations of the N–H bonds in the Si$_3$N$_4$ layer of the substrate [39, 40]. The intensity of this band decreases with the growth of the poly-Ge film thickness. The spectrum of the 200 nm thick poly-Ge film (Figure 16) has a feature in the form of deepening of the sum spectrum with the minimum at about 3800 cm$^{-1}$. It should be noted that the position and depth of this minimum depend on the growth temperature [11]. It occurs due to the effect of Ge film grain sizes and the surface roughness on the spectra as a contribution of the light scatter in the total light balance becomes significant. As mentioned above, with the increase in the film thickness, grains forming a polycrystalline layer coarsen, as a result their sizes as well as the surface roughness become comparable with the wavelength of the incident light in the short short-wavelength range; this causes intense light scattering, the contribution of which is not taken into account in the sum spectra. Notice that we have also observed a matte surface of this film in the visible light region that verifies the above statement.

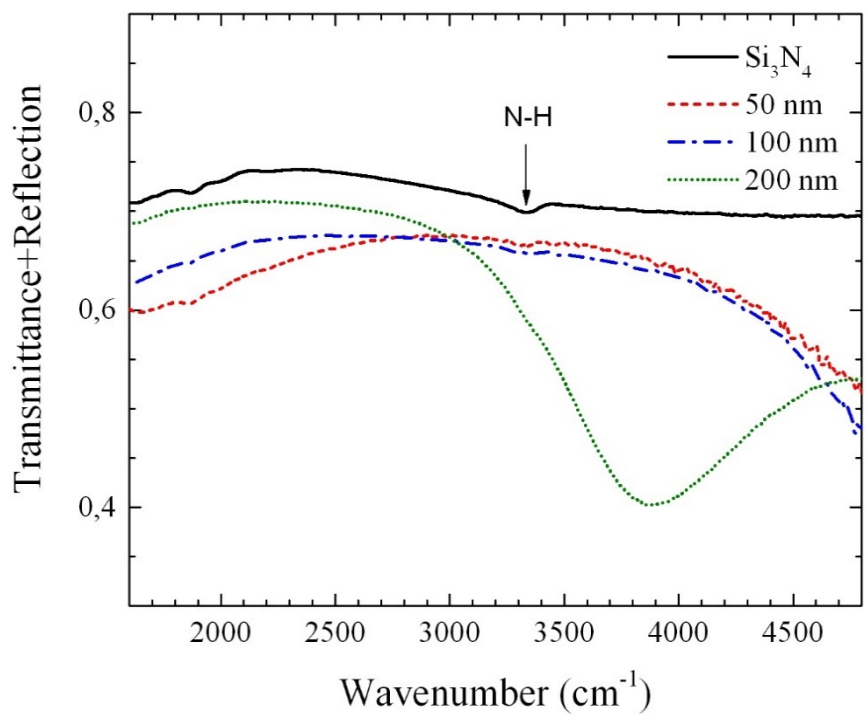

Figure 16. Sum of FTIR transmittance and reflectance spectra for the samples with poly-Ge films of different thicknesses grown at 500°C; the solid line relates to the Si$_3$N$_4$/SiO$_2$/Si(001) substrate.

## 4. Discussion

The deposition of the Si or Ge films on the Si$_3$N$_4$/SiO$_2$/Si(001) dielectric substrates led to modification of the IR absorption spectra detected as decreasing intensity of the N–H absorbance bands peaked at ~1150 cm$^{-1}$ and ~3340 cm$^{-1}$ (Figure 10, Figure 12 and Figure 16). In the Si$_3$N$_4$ layers of the Si$_3$N$_4$/SiO$_2$/Si(001) substrates, H atoms prefer forming bonds with N atoms because the N–H bond has a higher bonding energy (~ 4 eV) than the Si–H one has ( 3 eV) [42]. This is confirmed by the FTIR results since the IR absorption spectra of the Si$_3$N$_4$/SiO$_2$/Si(001) substrate do not show the band corresponding to the Si–H bond vibration usually placed around 2100 cm$^{-1}$ ( Figure 16), yet demonstrate the N–H absorption ones (Figure 10, Figure 11 and Figure 16) [39]. According to the FTIR results (Figure 10, Figure 12 and Figure 16), the increase in the film thickness essentially changed the intensity of the N–H absorption bands. The thicker was the grown film, the more considerable was the decrease in the N–H band intensity. This tendency was observed during growth of both amorphous and polycrystalline Si and Ge films. Based on the results of Raman spectroscopy, it should be noticed that the Si and Ge polycrystalline layers with different thickness had approximately equal magnitudes of the stress, with the internal stress being almost absent in inner parts of their grains: the corresponding Raman peak position did not differ from the that for the bulk single crystalline materials. Thus, the assumption that the layers with different thicknesses demonstrate a difference in strain, which is a reason of the observed dependence the hydrogen atom diffusion on the film thickness, is incorrect.

To explain of the obtained results, we employ the model of the hydrogen diffusion proposed in [43–49] hat was previously applied by us to the Si and Ge layer growth at different growth temperatures [10, 11]. In this work, the growth temperatures were not high enough in order to set up the favorable conditions of the H atom diffusion. The activation energy of the diffusion of hydrogen atoms in α-Ge and α-Si layers are ~ 1.5 eV [46 ,47, 50] and the bonding energies with N or Si atoms are ~4 eV and ~3eV, respectively. According to the model, the diffusion process can be described as movement of hydrogen atoms through interstitial states accompanied by the formation of the Si–H–Si (Ge–H–Ge) bonds. The Si–H (Ge–H) bonds, which emerge due to the capture of H atoms by dangling bonds of Si (Ge) atoms, are deep traps for hydrogen atoms. This process limits the diffusion. A base of the model is a concept of a hydrogen atom chemical potential ($\mu_H$). Hydrogen atoms bonding with deep traps (Si–H), due to higher bond energy, occupy a lower level than $\mu_H$. Hydrogen atoms migrating via interstitial states (week Si–Si (Ge–Ge) bonds), due to a lower bond energy, occupy a higher level than $\mu_H$ (the so-called transport level). If crystal structure has some disorder, then hydrogen atoms can transfer from $\mu_H$ to the transport level and easy diffuse. The $\mu_H$ level depends on the concentration of hydrogen atoms: the greater is concentration of hydrogen atoms, the higher is the $\mu_H$ level. In the structure formed by layers with different hydrogen content, a transfer of the hydrogen atoms from a part with a higher hydrogen atom content (a higher level of $\mu_H$) to one with a low hydrogen atom content (a low level of $\mu_H$) can be observed. The migration lasts until the chemical potential of hydrogen atoms becomes equal in both layers. This diffusion model originally brought forward for an amorphous layer is quite applicable to polycrystalline ones. Initially, hydrogen atoms are connected on the N–H bonds in the $Si_3N_4$ layer and diffusion can only begin when this energy barrier is overcome, e.g., due to a greater hydrogen atom content (a higher level of $\mu_H$) in the $Si_3N_4$ layer than in the growing film. A way of hydrogen atom diffusion in polycrystalline films is similar to that in amorphous ones (via interstitial states), and dangling silicon (germanium) bonds are deep traps for hydrogen, especially in the regions near grain boundaries [51].

In this study, the $Si_3N_4$ dielectric layers had a higher content of hydrogen atoms than others did and hydrogen could diffuse into the growing Si (Ge) films; this process activated already at the growth temperature as low as 30°C. Comparison of the IR absorbance spectra has shown that the higher the thickness of the deposited film was, the more hydrogen atoms left the $Si_3N_4$ layer during the growth of the amorphous and polycrystalline layers. This effect could be explained within the above model. During the film deposition, a region near the growth surface can be considered as a domain with the lowest hydrogen atom content. Thus, the gradient of the hydrogen atoms concentration (that of the chemical potential of hydrogen atoms) always exists in a growing layer and facilitates the transfer of hydrogen atoms from the enriched domain to the depleted one near the growth surface. As soon as the deposition process is stopped, the hydrogen atom diffusion is also finished. Probably, some content of hydrogen atoms remains in the Si (Ge) film, but since we have not observed the Si–H (Ge–H) absorption bands, either hydrogen atoms content was small or they formed such bonds with Si (Ge) atoms, which were inactive for the detection by FTIR.

The N–H band intensity decreased more noticeably in polycrystalline samples than in amorphous ones (for the 200 nm thick Ge films) that could be explained by higher growth temperatures (Figure 12). The increase in the growth temperature promotes the diffusion and desorption of hydrogen atoms from the growth surface and so the decrease in the hydrogen atoms content in the growing film [48, 52]. There was another feature observed for the thin 50-nm Ge films (Figure 12a, c). For the amorphous Ge layer, despite the low growth temperature, the intensity of the N–H absorption band decreased more significantly than for the polycrystalline one in comparison with value measured for the $Si_3N_4/SiO_2/Si(001)$ substrate. At the same time, the comparison of changes in the Si–N absorption bands of the thin 50 nm poly-Ge (Figure 15 a) and α-Ge films (Figure 15 b) also reveals that the band intensity decreases with the film thickness faster during the deposition of the α-Ge layer than the poly-Ge one. We suppose that the observed changes may be related to different content of the Ge dangling bonds in the growing polycrystalline and amorphous films. At the initial growth stage, polycrystalline films are shown to be formed by fine grains and to have a large area of grain boundaries, which are characterized by a large number of dangling bonds slowing down the diffusion of H atoms. As the film thickness increases, the grains become coarser, and the boundaries decrease in their area. Thus, the influence of dangling bonds could be more pronounced during the growth of thinner polycrystalline films than thicker ones. Now we cannot explain how a high content of dangling bonds influences the H atom diffusion. We consider two mechanisms each of which affects the $\mu_H$ level in the region of the growing film near the interface with the $Si_3N_4$ layer. First, being deep traps the dangling bonds could capture H atoms and change the $\mu_H$ level by increasing the content of H atoms in the growing film near the interface. However, in the IR spectra, we have not observed absorption bands corresponding to the Si–H and Ge–H bonds neither for the polycrystalline nor for the amorphous films. Second, the high concentration of dangling bonds themselves may affect the $\mu_H$ level and slow down the transfer of H atoms from the $Si_3N_4$ layer into the growing film.

The deposition of the Si(Ge) films also changed the strength of the Si–N and Si–O (~960 cm$^{-1}$) absorption bands in the IR spectra (Figure 10, Figure 12, Figure 13 and Figure 15). In the range from 650 to 1350 cm$^{-1}$, the increase of the Si–N peak intensity is connected with the formation of additional Si–N bonds in the $Si_3N_4$ layer. After the breaking of the N–H bonds and the outmigration of hydrogen atom into the growing Si (Ge) film, a dangling bond of a nitrogen atom could be closed by forming a bond with the nearest Si atom. Si (Ge) atoms entering the $Si_3N_4$ layers from the growing film have an additional effect on the intensity. These atoms connect to dangling bonds of N atoms to form the Si–N (Ge–N) bonds, which increase (decreases) the strength of the Si–N absorption band near 860 cm$^{-1}$. This process was reviewed in detail in our previous articles [10, 11]. The formation of the Si–N (Ge–N) bonds improves order in the atom arrangement in the Si$_3$N$_4$ layers and leads to the decrease in the intensity of the Si–N absorption band near ~500 cm$^{-1}$ [41]. The changes in the intensity of the Si–O band (~960 cm$^{-1}$) are connected with the process of the Si (Ge) atom diffusion though the top oxynitride layer in the Si$_3$N$_4$ one, which takes place simultaneously with the hydrogen atom diffusion [11]. A higher growth temperature and a thicker film facilitate the decrease in the strength of the Si–O absorption band near 960 cm$^{-1}$. This process is the most noticeable during the germanium film growth.

## 5. Conclusion

The results of FTIR and Raman spectroscopy studies of silicon and germanium films with the thickness from 50 to 200 nm deposited on a dielectric Si$_3$N$_4$/SiO$_2$/Si(001) substrate are presented in this work. A considerable reduction of the strength of the N–H absorption bands in the IR spectra has been observed as a result the growth of the silicon (germanium) films; the thicker the film was grown, the more significant the decrease in the intensity of the N–H band. This trend was observed during the depositing of both the amorphous and polycrystalline Si and Ge films. According to Raman spectroscopy results, the main band in the spectra related to the samples with the polycrystalline silicon films was located around 520 cm$^{-1}$ and decomposed to three components peaked at ~519 cm$^{-1}$, ~518 cm$^{-1}$ and ~500 cm$^{-1}$. The first of them relates to the TO(c-Si) and corresponds to the inner crystalline region of the grains. The latter two peaks are due to Raman scattering at the grain boundaries of polycrystalline layer. The main band in the spectra of the samples with the germanium films reached maximum at ~300 cm$^{-1}$ and consisted of two peaks at ~300 cm$^{-1}$ and ~290 cm$^{-1}$, which were connected with the TO(c-Ge) phonons of the internal crystalline part of the grains and grain boundaries of the polycrystalline film, respectively. An increase in the thickness of a grown film led to significant shifts in peak positions in the Raman spectra for neither the silicon layers nor the germanium ones.

The presented experimental results may be explained using a model proposed in Ref. [43]. The main point of model is that the diffusion of hydrogen atoms is controlled by the difference in chemical potentials of hydrogen atoms in the dielectric $Si_3N_4$ layer and the growing silicon or germanium film. The difference in chemical potentials of hydrogen atoms in the dielectric $Si_3N_4$ layer and the growing film remains until the film growth is stopped and the film thickness ceases increasing. Therefore, a nonstop increase in the film thickness supports the diffusion of hydrogen atoms from the dielectric $Si_3N_4$ layer into the growing Si or Ge layer until the exhaustion of hydrogen atoms in the dielectric layer.

**Acknowledgments**

This research did not receive any specific grant from funding agencies in public, commercial or not-for-profit sectors.

The Center for Collective Use of Scientific Equipment of GPI RAS supported this research via presenting admittance to its equipment.

**Conflicts of interest**

No conflicts of interest exist, which could potentially influence the work.

**CRediT author statement**

| | |
|---|---|
| Larisa V. Arapkina: | Investigation, Data Curation, Writing - Original Draft |
| Kirill V. Chizh: | Investigation, Data Curation, Writing - Original Draft |
| Mikhail S. Storozhevykh: | Investigation, Data Curation, Writing - Original Draft |
| Dmitry B. Stavrovsky: | Investigation |
| Alexey A. Klimenko: | Investigation |
| Alexander A. Dudin: | Investigation, Data Curation |
| Vladimir P. Dubkov: | Investigation |
| Vladimir A. Yuryev: | Data Curation, Writing - Review & Editing, Supervision |